\documentstyle[preprint,tighten,prd,aps,psfig,floats]{revtex}

\def\mpl{m_{\rm Pl}}

\def\ombh2{\Omega_{\rm B}h^2}
\def\om0{\Omega_0}

\def\sig8{\sigma_8}

\def\beq{\begin{equation}}
\def\eeq{\end{equation}}

\def\ie{{\it i.e.}\ }

\begin{document}
\title{Gravity waves goodbye}
\author{J. P. Zibin and Douglas Scott}
\address{Department of Physics and Astronomy, 
University of British Columbia, 6224 Agricultural Road, 
Vancouver, BC V6T 1Z1 Canada}
\author{Martin White}
\address{Departments of Astronomy and Physics, University of Illinois at
Urbana-Champaign, 1110 West Green Street, Urbana, IL 61801-3080}
\date{\today}
\maketitle

\begin{abstract}

\noindent The detection of a stochastic background of long-wavelength
gravitational waves (tensors) in the cosmic microwave background (CMB)
anisotropy would be an invaluable
probe of the high energy physics of the early universe.  Unfortunately a
combination of factors now makes such a detection seem unlikely:
the vast majority of the CMB signal appears to come from density perturbations
(scalars) - detailed fits to current observations indicate a
tensor-to-scalar quadrupole ratio of $T/S<0.5$ for the
simplest models; and on the theoretical side the best-motivated
inflationary models seem to require very small $T/S$.
Unfortunately CMB temperature anisotropies can only probe a gravity wave
signal down to $T/S\sim 10\%$ and optimistic assumptions about polarization
of the CMB only lower this another order of magnitude.

\end{abstract}

\bigskip
\bigskip
\bigskip

Inflation is the only known mechanism for producing an almost scale-invariant
spectrum of adiabatic scalar (density) fluctuations, a prediction which is
steadily gaining observational support.
The simplest models of inflation also predict an almost
scale-invariant spectrum of gravity waves.
A definitive detection of these waves would constitute a window onto physics
at higher energies than have ever been probed before.  Indeed, it has been
realized for some time that, for monomial inflation models within the slow-roll
approximation, measurement of this spectrum could allow a reconstruction of
the inflaton potential itself \cite{InfReview}.
Unfortunately, a combination of factors now makes this seem unlikely:
the vast majority of the CMB signal probably comes from scalar perturbations,
for reasons we shall now describe.

Theoretically one could imagine that whatever mechanism produces the initial
fluctuations which seeded structure formation would produce all three types of
perturbations (scalar, vector and tensor) roughly equally.  If this happens
early enough, the vector modes -- representing fluid vorticity -- would decay
with the expansion of the universe leaving only scalar and tensor perturbations
today.  Unfortunately we live in a ``special'' universe in which the
perturbations appear to be both adiabatic and close to scale-invariant
(equal contribution to the metric perturbation per logarithmic interval in
wavelength).  Our best paradigm for producing such fluctuations is
amplification of quantum fluctuations by a period of accelerated expansion,
\ie inflation.  Within this paradigm we shall see that $T/S$ is expected
to be considerably smaller than the naive $1:1$ ratio.

While some simple models of inflation predict $T/S\simeq 1$, we would hope that
inflation would one day find a home in modern particle physics theories.
{}From this perspective, there is currently a ``considerable theoretical
prejudice against the likelihood'' \cite{lyth97} of an observable gravity
wave signal in the CMB anisotropy.
While our knowledge of physics above the electroweak scale is extremely
uncertain, a large $T/S$ requires two unlikely events.
First the scale of variation of the inflaton field during inflation would
need to be ${\cal O}(\mpl)$ or greater, which is inconsistent with an ordinary
extension of the standard model \cite{lyth97}.
Secondly, the size of the inflaton potential would necessarily be
$V^{1/4}\sim 10^{-2}\mpl$, orders of magnitude larger than generically
expected from particle theory \cite{lyth97}.
Thus inflation, in a particle physics context, predicts that the scalar
perturbations will be dramatically enhanced over the tensor perturbations.
A separate line of argument, involving a description of inflation
motivated by quantum gravity, leads to a prediction of
$T/S \simeq 1.7 \times 10^{-3}$ \cite{awt}.

What is the situation on the observational side?
The measurement of the large-angle CMB anisotropies by the {\sl COBE\/}
satellite has been followed by ground- and balloon-based observations
of the smaller scale regions of the CMB power spectrum.  Measurements
on a range of scales are needed to constrain gravity waves,
since tensors are expected to contribute only to angular scales greater
than about $1^\circ$.  Thus large-scale power greater than that expected
from the extrapolation of small-scale scalar power can be attributed
to tensors.  This small-scale power can be measured from the CMB itself
or additionally from matter fluctuations in more recent epochs, allowing
a large lever arm in scale.

We have recently placed upper limits on $T/S$ using a variety of observations
\cite{zsw}.  We used CMB anisotropy data as well as
information about the matter fluctuations from galaxy correlation,
cluster abundance, and Lyman~$\alpha$ forest measurements.
Our limits include a variety of additional constraints (such as the age of the
universe, cluster baryon density, and recent supernova measurements), in all
cases marginalizing over the relevant but as yet imprecisely
determined cosmological parameters.
We placed constraints on exponential and polynomial inflaton potential models;
these ``large-field'' models predict substantial
 $T/S$ and are therefore of interest here.
We found $T/S<0.5$ at 95\% confidence, with the small-angle CMB
data providing the bulk of the constraint (see Fig. 1).

\begin{figure}
\centerline{\psfig{figure=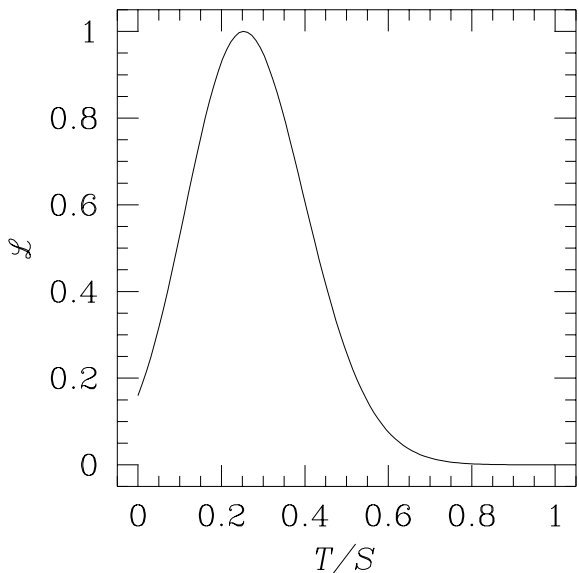,height=8cm}}
\caption{Likelihood versus tensor-to-scalar ratio for exponential inflaton
potentials, using all the CMB and matter power spectrum observations
mentioned in the text.  Note that the results are fully consistent with
$T/S = 0$.}
\end{figure}

In the next several years a pair of satellite missions should dramatically
improve our picture of the CMB.  {\sl MAP\/} and especially
the {\sl Planck\/} Surveyor will map the CMB at unprecedented
precision.  What can we expect from these missions regarding gravity
wave constraints?  With a cosmic variance limited experiment capable
of determining only the anisotropy in the CMB, but with all other
parameters known, one can measure $T/S$ only if it is larger than about
10\% \cite{KnoTur}.  To additionally measure the tensor spectral index
and check the inflationary consistency relation \cite{InfReview}
requires $T/S$ to be a factor of several larger.  This is in conflict
with current theoretical prejudice, and realistically also in conflict
with the experimental limits.

One can potentially improve sensitivity to $T/S$ by measuring the polarization
of the CMB.  With more observables the error bars on parameters are
tightened.  In addition, polarization breaks the degeneracy between
reionization and a tensor component, both of which affect the relative
amplitude of large and small angular scales, allowing extraction of smaller
levels of signal \cite{ZalSel}.  Model dependent constraints on a
tensor perturbation mode as low as $1\%$ appear to
be possible with the {\it Planck\/} satellite
\cite{constr}, though numerical inaccuracies plagued
earlier work \cite{EHT} making these numbers somewhat soft.

Scalar modes have no ``handedness'' and hence they generate only parity even,
or $E$-mode polarization \cite{ZalSel,KKS}.  A detection of
$B$-mode polarization would thus indicate the presence of other modes, with
tensors being more likely since vector modes decay cosmologically.
Unfortunately the detection of a $B$-mode polarization will be a formidable
experimental challenge.  The level of the signal is expected to be very small:
only a few tens of nK.
One can regain some signal-to-noise ratio by concentrating on the sign of the
correlation of polarization with temperature anisotropies on large-angular
scales (scalar polarization is tangential around hot spots, while tensor
polarization is radial).
Unfortunately only a small fraction of the signal is correlated, so again
the signal is extremely small (less than $1\mu$K), and the correlation is
swamped by the scalar signal unless $T/S$ is significant.

The ground-based laser interferometers LIGO and VIRGO, the
proposed space-based interferometer LISA,
and millisecond pulsar timing offer another conceivable route to primordial
gravity wave detection.  However, the long lever arm from the horizon scale
to the scales probed by these experiments makes direct detection infeasible
\cite{kw,TurWhi,CalKamWad}.

All of these arguments combine to considerably reduce the optimism for
the detailed reconstruction of the inflaton potential through
cosmological gravity wave measurement.  However, since inflation
appears to predict low $T/S$, it is good news that observations
support a small tensor contribution.  The apparent
demise of the significance of tensors for the CMB has one important
consequence:  detection of even a modest contribution of gravity waves
would profoundly affect our view of early universe particle physics.


\acknowledgements
This research was supported by the Natural Sciences and Engineering 
Research Council of Canada and by the U. S. National Science Foundation.


\begin{thebibliography}{10}

\bibitem{InfReview}
A. R.~Liddle and D. H.~Lyth, Phys. Rep. {\bf 231}, 1 (1992);
J. E.~Lidsey, et al., Rev. Mod. Phys. 69, 373 (1997);
D. H.~Lyth and A. Riotto, Phys. Rep., in press, [hep-th/9807278].

\bibitem{lyth97}
D.~H.~Lyth, Phys. Rev. Lett. {\bf 78},  1861  (1997);
D.~H.~Lyth and A. Riotto, Phys. Rep., in press, [hep-th/9807278].

\bibitem{awt}
L. R. Abramo, R. P. Woodard, and N. C. Tsamis, preprint [astro-ph/9803172].

\bibitem{zsw}
J. P.~Zibin, D.~Scott, and M.~White, preprint [astro-ph/9901028].

\bibitem{KnoTur}
L.~Knox and M.~S. Turner, Phys. Rev. Lett. {\bf 73}, 3347 (1994);
L.~Knox, Phys. Rev. D {\bf 52}, 4307 (1995).

\bibitem{ZalSel}
M. Zaldarriaga and U. Seljak, Phys. Rev. D {\bf 58}, 023003 (1998).

\bibitem{constr}
M. Zaldarriaga, D. N. Spergel, and U. Seljak, Astrophys. J., {\bf 488},
1 (1997);  J.~R. Bond, G. Efstathiou, and M. Tegmark, Mon. Not. R. Astron.
Soc., {\bf 291}, L33 (1997);
M. Kamionkowski and A. Kosowsky, Phys. Rev. D {\bf 57}, 685 (1998);
W.~H. Kinney, preprint [astro-ph/9806259].

\bibitem{EHT}
D. J. Eisenstein, W. Hu, and  M. Tegmark, preprint [astro-ph/9807130].

\bibitem{KKS}
M. Kamionkowski, A. Kosowsky, and A. Stebbins, Phys. Rev. D {\bf 55},
7368 (1997).

\bibitem{kw}
L.~M. Krauss and M. White, Phys. Rev. Lett. {\bf 69},  869  (1992).

\bibitem{TurWhi}
M.~S. Turner and M.~White, Phys. Rev. D {\bf 53}, 6822 (1996).

\bibitem{CalKamWad}
R. R.~Caldwell, M.~Kamionkowski, and L.~Wadley, preprint [astro-ph/9807319].

\end{thebibliography}
\end{document}